\documentclass[11pt]{article}
\usepackage{moriond,epsfig}

\def\Journal#1#2#3#4{{#1} {\bf #2}, #3 (#4)}

\def\PRP{{\em Phys. Rep. }}  
\def\NIMA{{\em Nucl. Instrum. Methods} A}
\def\NPB{{\em Nucl. Phys.} B}
\def\PLB{{\em Phys. Lett.}  B}

\def\PRD{{\em Phys. Rev.} D}

\def\ra{\rightarrow}

\def\be{\begin{equation}}
\def\ee{\end{equation}}
\def\bea{\begin{eqnarray}}
\def\eea{\end{eqnarray}}

\newcommand{\exps}{experiments } 
\newcommand{\evs}{\mbox{eV$^2$} } 
    
\newcommand{\bnmu}{\mbox{$\bar{\nu}_\mu$} }    
  
\newcommand{\nel}{\mbox{$\nu_e$} }   
\newcommand{\nmu}{\mbox{$\nu_\mu$} }  
\newcommand{\ntau}{\mbox{$\nu_\tau$} }     
\newcommand{\sint}{\mbox{$\sin^2 2\theta$} }   
\newcommand{\osz}{oscillation } 
 
\newcommand{\neu}{neutrino }  
   
\newcommand{\nmuntau}{\mbox{$\nu_{\mu} - \nu_{\tau}$} } 
\newcommand{\nentau}{\mbox{$\nu_e - \nu_{\tau}$} } 
\newcommand{\nmune}{\mbox{$\nu_{\mu} - \nu_e$} } 
\newcommand{\delm}{\mbox{$\Delta m^2$} }   
\newcommand{\dnull}{\mbox{D$^0$} } 
\newcommand{\ds}{\mbox{D$^{\ast +}$} } 
\newcommand{\ch}{CHORUS }
\newcommand{\no}{NOMAD }
\newcommand{\enu}{\mbox{$E_{\nu}$} }        
\begin{document}
\vspace*{4cm}
\title{LATEST CHORUS AND NOMAD RESULTS}

\author{ K. Zuber }

\address{Department of Physics, University of Oxford, Keble Road,\\
Oxford OX1 3RH, England}

\maketitle\abstracts{
The final result of the NOMAD searches on \nmu - \ntau oscillations as well as the current status of 
CHORUS are described. The \nmune analysis of \no excludes 
the parameter region of LSND in the range \delm $>10$ \evs. New 
results on charm physics from both experiments are presented.}

\section{Introduction}
A non-vanishing rest mass ot the \neu has far reaching consequences from cosmology down to particle 
physics. For a recent review see \cite{kai}. 
In the last years growing evidence for such a mass arose in \neu \osz \exps .  
In a simple two flavour mixing scheme the \osz probability $P$ is given by
\be  
P (L/E_\nu) = \sint{} \sin^2(1.27 \delm_{ij} L/E_\nu)  
\ee  
with $\Delta m^2_{ij}$ = $\mid m_j^2 -m_i^2 \mid$, $\theta$ as the mixing angle, L the
source-detector distance and \enu the 
\neu energy. Short baseline experiments with high energy neutrino beams offer the chance
to probe small mixing angles at rather larger \delm (\delm $>$ 1 \evs). It was the intension of
\ch and \no to probe \sint almost down to about $10^{-4}$ for \nmuntau oscillations 
in that region. High statistics $\nu$N scattering allows also for various other kinds of
studies,
among them charm physics.

\section{The experiments}
\label{subsec:prod} 
Both experiments were performed in the West Area Neutrino Facility (WANF)
at CERN during the years 1994-1998. The beam was an almost pure \nmu beam with small contaminations of \bnmu
($\approx$ 6 \%) and \nel ($\approx$ 1 \%).  
The active target of \ch consisted of nuclear emulsions with a total mass of 770
kg \cite{chorus}. 
For timing purposes and for extrapolating the tracks back into the emulsions, a scintillating fibre tracker was
interleaved.
Behind the target complex followed a hexagonal spectrometer
magnet for
momentum measurement, a high resolution spaghetti calorimeter for measuring hadronic showers and a muon spectrometer.
The scanning of the emulsions is performed with high-speed CCD microscopes.\\
\no was using drift chambers as target and tracking medium \cite{nomad}.
In total there were 44 chambers located in a 0.4 T magnetic field with a fiducial mass of 2.7 tons.
They were followed by a transition radiation detector (TRD) for e/$\pi$ separation,
further electron identification devices in form of a preshower detector
and an electromagnetic lead glass calorimeter. 
A hadronic calorimeter 
and a set of 10 drift chambers for muon identification followed. 
In front of the drift chambers another iron-scintillator calorimeter of about 20 t target mass was
installed. 

\section{\nmuntau and \nentau analysis}
\begin{center}
\begin{figure}[ht]
\hskip 2cm
\epsfig{figure=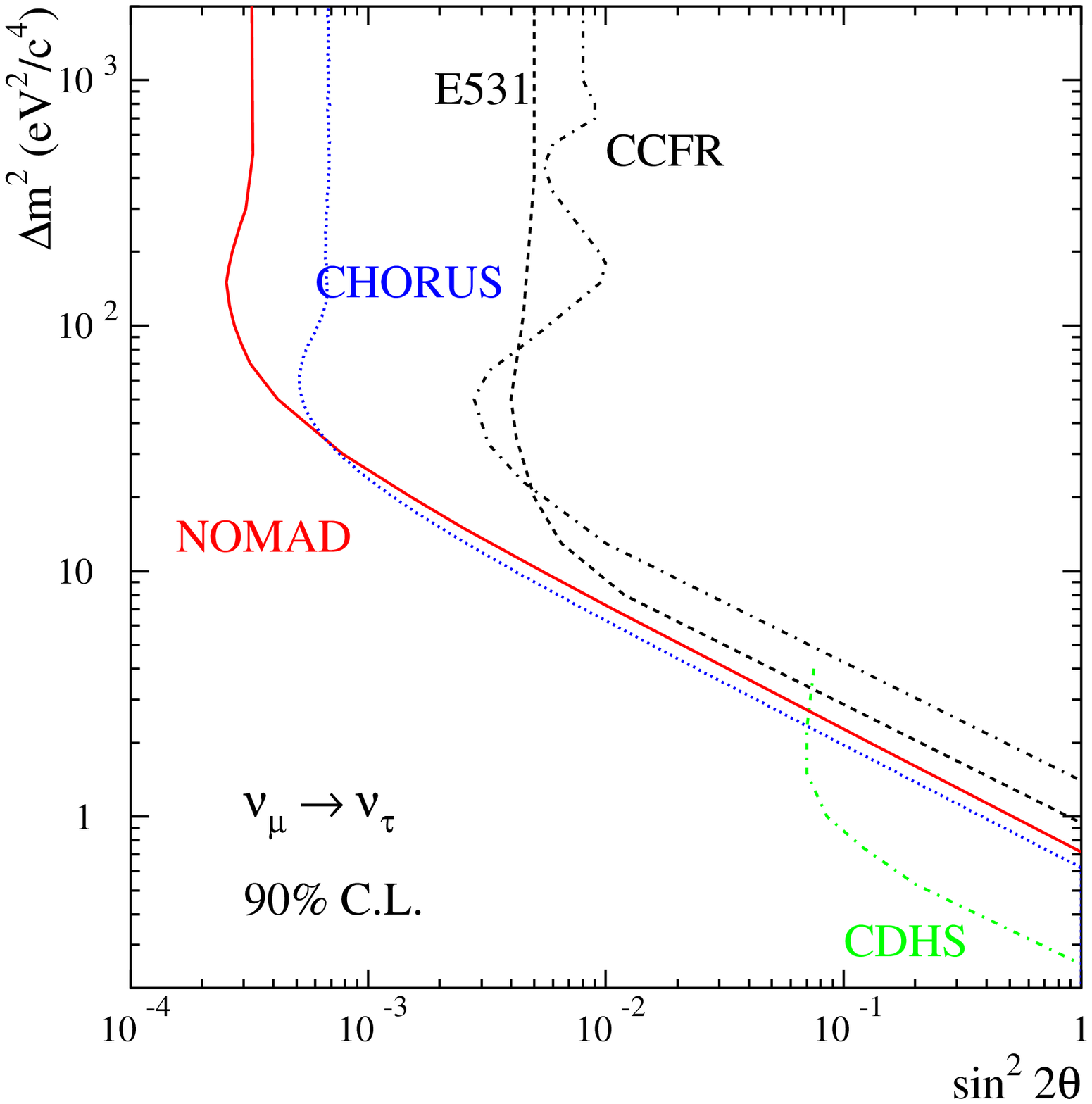,width=5.5cm,height=5.5cm} 
\hskip 1cm
\epsfig{figure=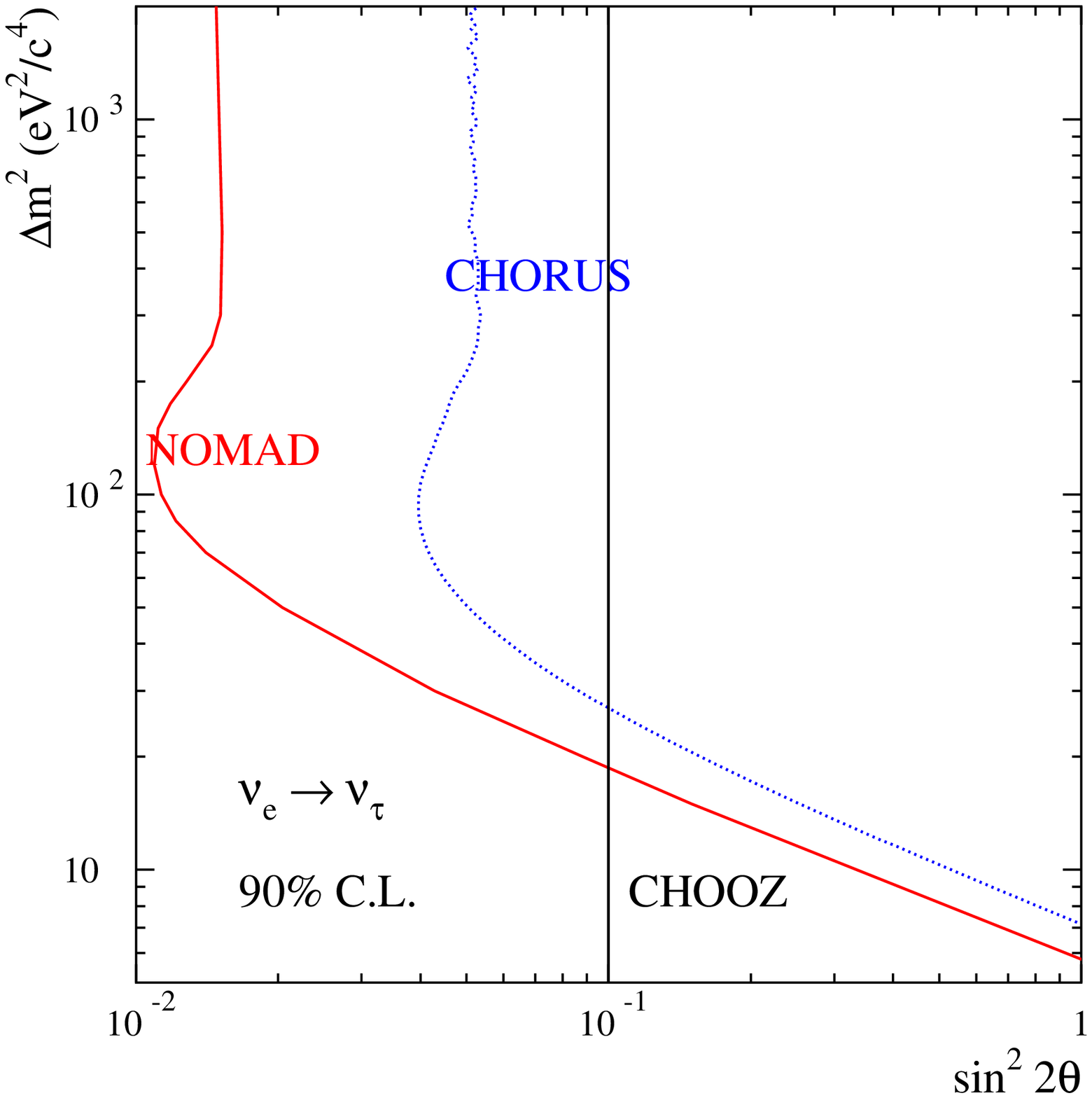,width=5.5cm,height=5.5cm} 
\caption{Current status of \nmuntau (left) and \nentau (right) oscillation searches obtained by \ch and \no.
Superimposed are results from other experiments \label{fig:osci}}  
\end{figure}  
\end{center}

The analysis obtained by \ch is based on the search of a kink, coming from the decay of the $\tau$-lepton.
The
data set are divided in 1$\mu$ and 0$\mu$ samples based on the possible observation of a muon in the spectrometer. To
reduce the load for scanning, further cuts on momentum and angles with respect to the beam are applied. 
The event finding proceeds in three steps: a vertex location in the emulsion, the search for the
decay kink and at the end an eye-scan of candidates. The main background arises from charm production.
\ch observed 0 events in the 1$\mu$ (0.1 expected)
and 4 events in the 0$\mu$ sample (3.3 expected). This can be transformed into an upper limit on
the \osz probability of P(\nmuntau) $< 3.4 \cdot 10^{-4}$ (90 \% CL) \cite{chnutau}. With a new, more effective scanning
method the
analysis will be repeated.\\
\no is not able to observe the $\tau$-lepton directly but relies on the decay kinematics. The variables
of use here are basically imbalance and isolation. 
To obtain maximum sensitivity for an oscillation signal all degree of freedoms of the kinematics and their
correlations have to be exploited. In this way the $\tau \ra e$ and $\tau \ra h(n \pi^0), 3h(n\pi^0)$ decay
modes were explored, separated in a low multiplicity and deep inelastic sample. 
A maximum likelihood method was used
together with a blind-box analysis. The box itself is split
into bins, where most of the sensitivity is related to low background bins. 
In these bins 1.60$^{+1.69}_{-.39}$ background events were expected and 1 event was observed. The final result from the
analysis is a limit of P(\nmuntau) $< 1.68 \cdot 10^{-4}$ (90 \% CL) \cite{nonutau}.\\
Both experiments took advantage of the fact, that there is a 1\% beam contamination of $\nu_e$. Correcting for
the higher average energy of the \nel component and modified acceptances, they were able to give
limits also on \nentau oscillation probabilities as P(\nentau) $< 2.6 \cdot 10^{-2}$ (\ch ) and P(\nentau) $< 1.68 \cdot
10^{-2}$ (\no ) respectively. The current status is shown in Fig.~\ref{fig:osci}.

\section{\nmune analysis}
\begin{figure}[hb]
\hskip 2cm
\epsfig{figure=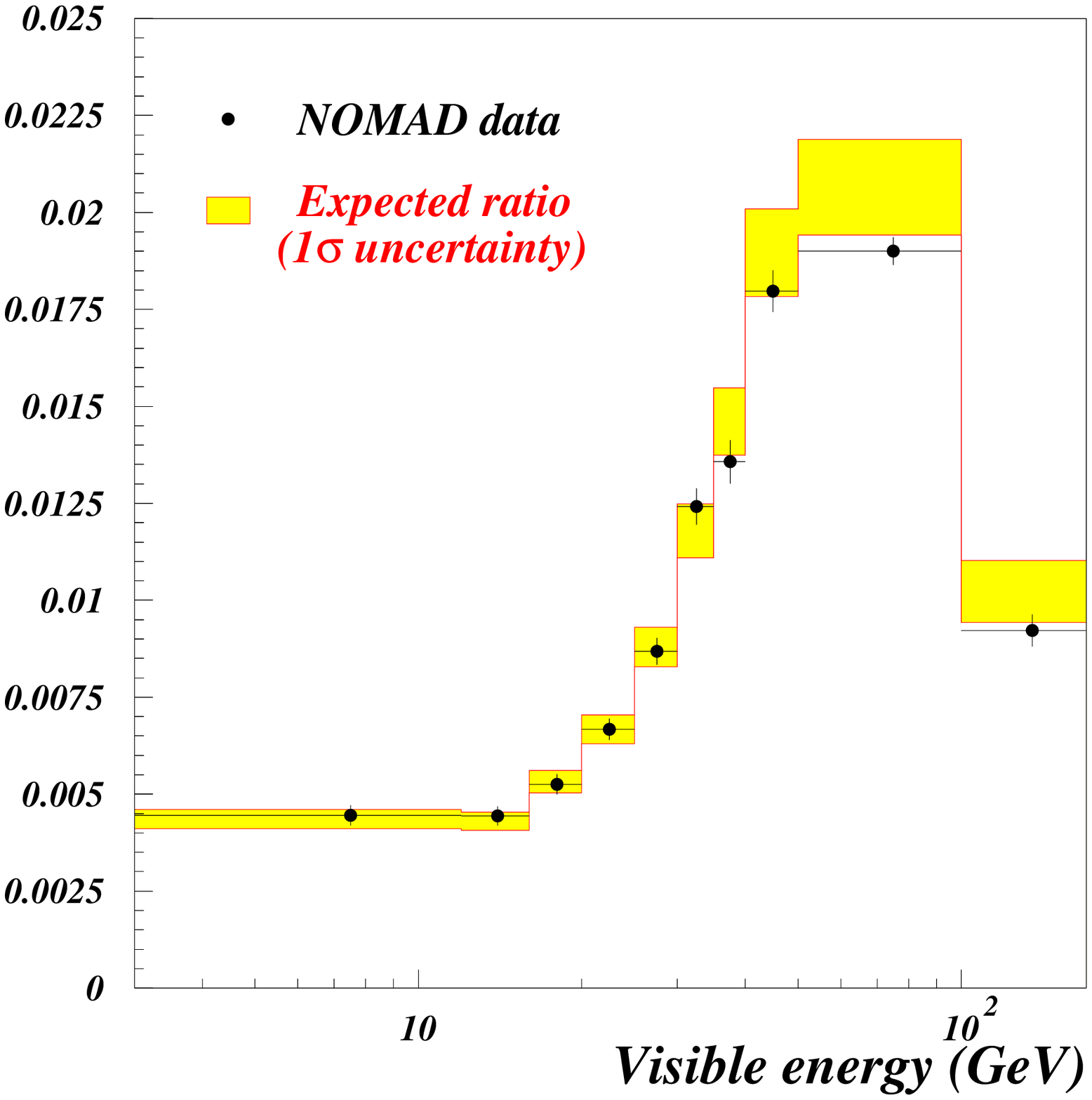,width=5cm,height=5cm} 
\hskip 1cm
\epsfig{figure=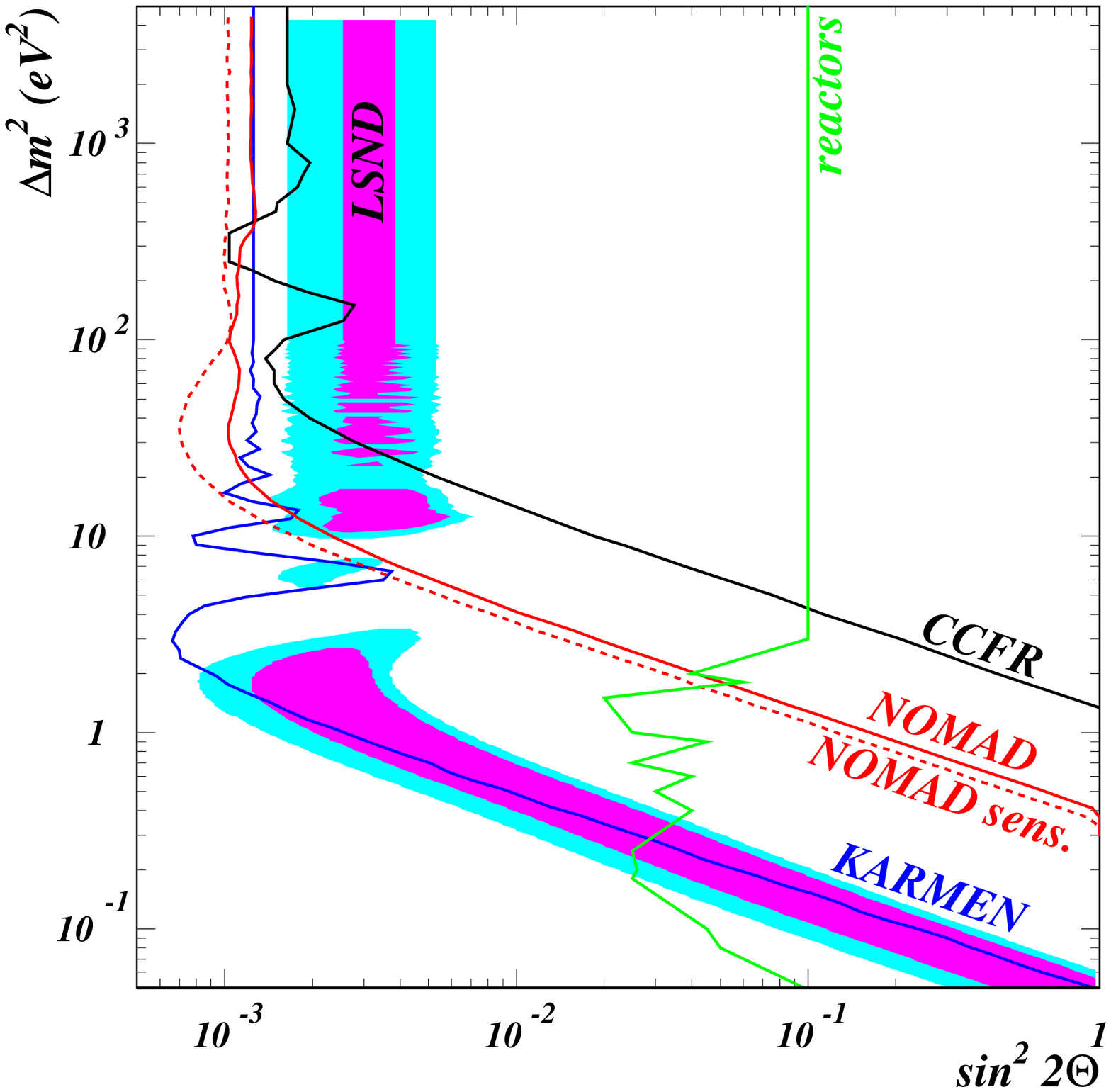,width=5cm,height=5cm} 
\caption{Left: $R_{e\mu}$ as a function of visible energy. The points show the \no data, the histogram
corresponds to the expected Monte Carlo predictions assuming no oscillations and 1$\sigma$ systematic errors
added in quadrature. Right: 90 \% CL exclusion region in the \delm - \sint plot and the sensitivity of the
\no analysis superimposed on the results of other experiments. A new KARMEN analysis\protect \cite{karmen} and a
combined
LSND-KARMEN analysis\protect \cite{klaus} exist.
\label{fig:nmune}}  
\end{figure}

A \nmune signal would manifest itself in \no as an increase of \nel CC events. The difference in the energy and radial
distributions of incident \nel and \nmu lead to an enhancement of \nel CC events at low \nel energies and small radii
with respect to
the beam
axis. To reduce systematic effects the ratio $R_{e\mu}$ defined as
\be
R_{e\mu} = \frac{'' \nel CC''}{'' \nmu CC''} (\enu, r) 
= \frac{e^- (\nel CC) + e^- (bkg) + e^- (\nel^{osc} CC)}{\mu^- (\nmu CC) + \mu^- (bkg)} (\enu, r) 
\ee
is investigated.
The sensitivity can be enhanced to explore $R_{e\mu}$ as a function of neutrino energy \enu and
the radial position of the neutrino interaction vertex $r$. The analysis is also based on kinematical 
criteria and performed as a ''blind'' analysis.  The crucial part and main uncertainty is a robust
prediction of the neutrino flux and energy spectrum.
No oscillations were observed \cite{slava} resulting in a value of \delm $>$ 0.4 \evs for maximal mixing and
\sint $< 1.2 \cdot 10^{-3}$ for large $\Delta m^2$. \no excludes the LSND evidence\cite{lsnd} for \delm
$>$ 10 \evs. 
The $R_{e\mu}$ ratio and the resulting exclusion plot
are shown in Fig.~\ref{fig:nmune}.

\section{Charm physics} 
\begin{figure} 
\hskip 2cm
\epsfig{figure=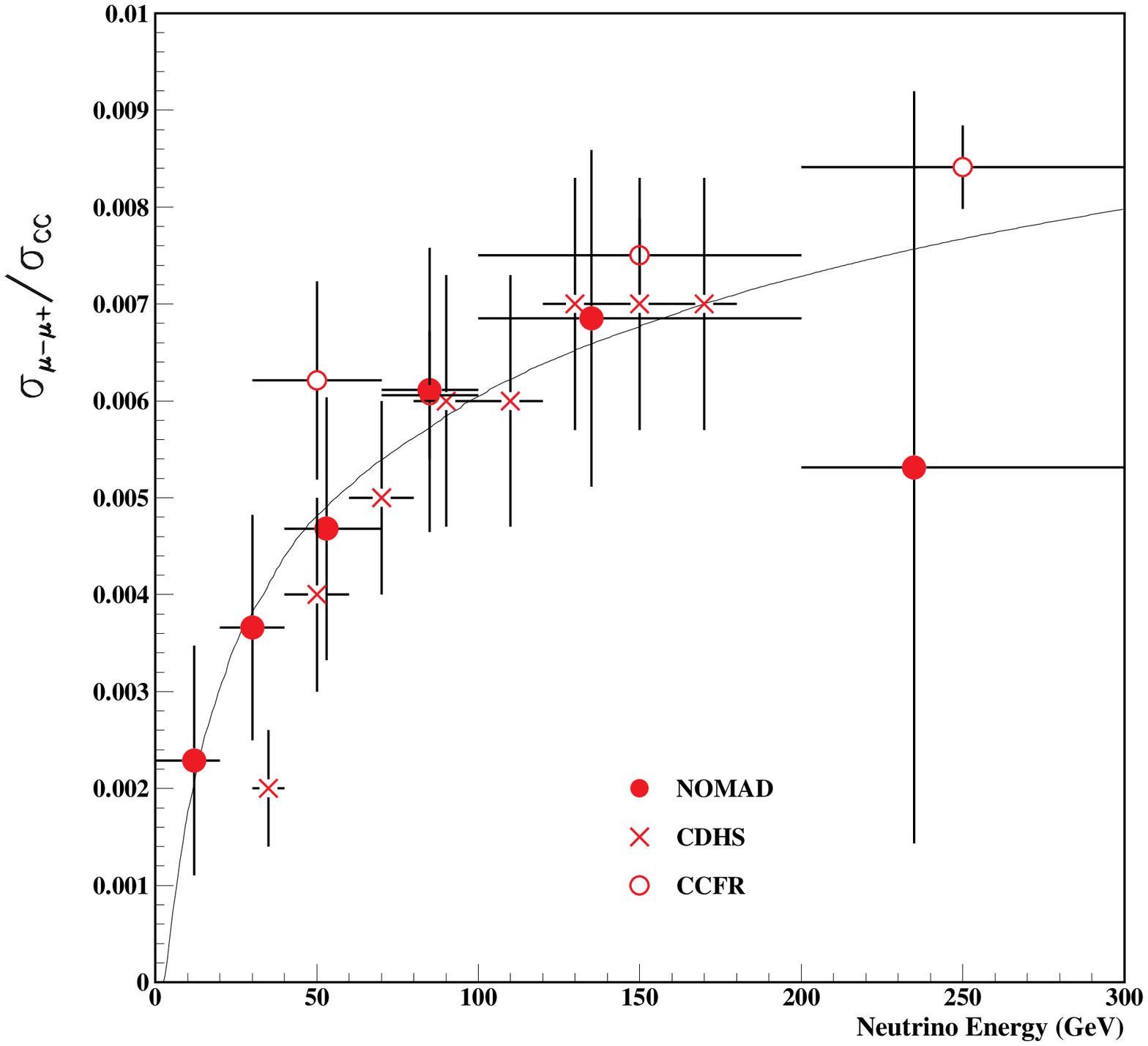,width=5.cm,height=5.5cm}
\hskip 1cm
\epsfig{figure=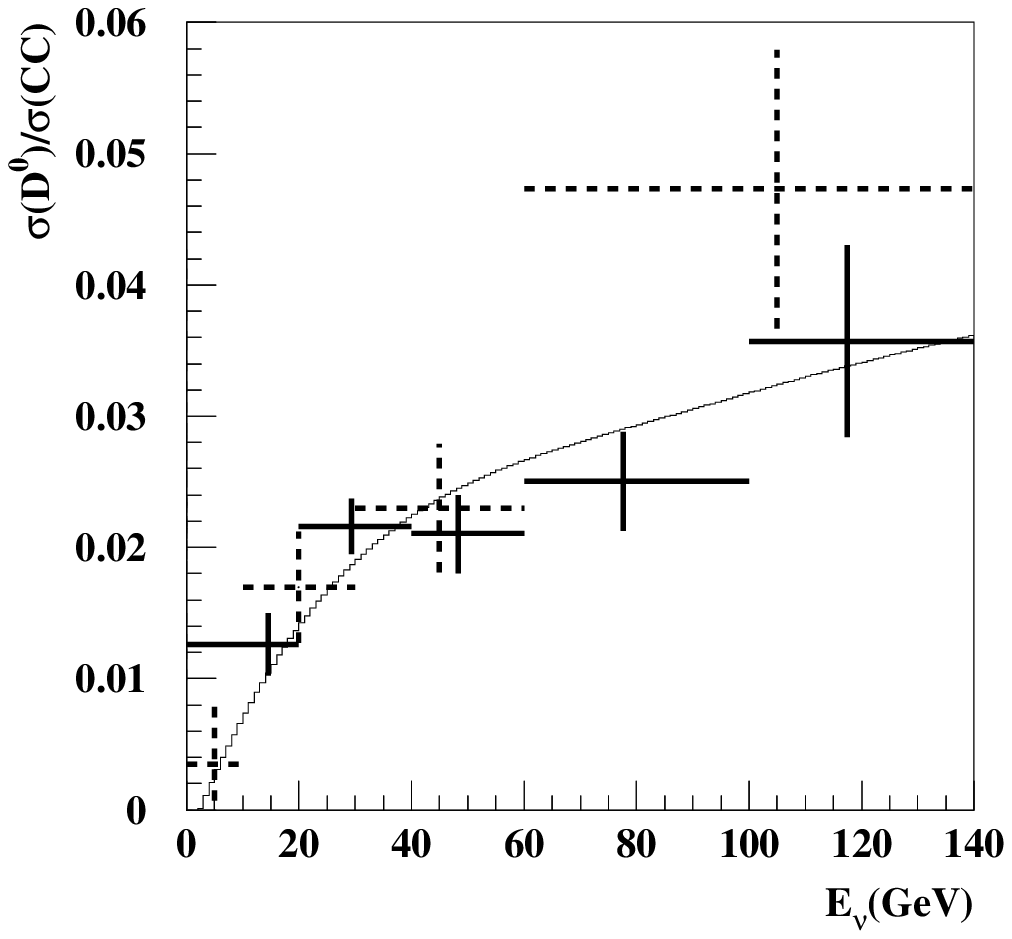,width=5.5cm,height=5.5cm} 
\caption{Left: $\sigma(\mu^+\mu^-)/\sigma(\nmu CC)$ as a function of \enu for
a charm quark mass of $m_c = 1.3 GeV/c^2$. Also shown are the CCFR
(open circles) and CDHS data (crosses). The theoretical curve is using the slow rescaling model in LO
QCD, a charm quark mass of $m_c = 1.3 GeV/c^2$, a strange quark suppression of $\kappa = 0.48$, an average semileptonic
branching ratio of 0.095 and the validity of the CCFR structure functions.
Right:  \dnull production rate as a function of \enu as obtained by \ch. The result of this analysis are shown as solid
lines and
compared with those of the E531 experiment, scaled appropriately (dashed lines). 
The curve shows a fit based on the slow
rescaling model to NOMAD charm data multiplied by the (\dnull/charm) cross-section ratio measured.
\label{fig:charm}} 
\end{figure}

Charm can be produced via charged current interactions on d- and s-quarks in the nucleon, therefore allowing to explore
their structure functions in the nucleon, especially s(x). 
The charm fragmention can be explored by measuring the various charmed mesons in the final state. 
From the ratio of opposite sign dimuons versus single muon production as a function of \enu,
$m_c$ can be determined from the threshold behaviour. Such an analysis\cite{noosdm} was performed by \no using 
15 \% of the data set (Fig.~\ref{fig:charm}). 
Recently \ch performed a search for \dnull production \cite{chdo}. In total 283 candidates are observed, with an
expected
background of 9.2 events coming from K and $\Lambda$ decay. The ratio $\sigma(\dnull)/\sigma(\nmu CC)$
is found to be $(1.99 \pm 0.13(stat.)\pm0.17(syst.))\cdot 10^{-2}$ at 27 GeV average \nmu energy
(Fig.~\ref{fig:charm}).
NOMAD performed a search for \ds -production using the decay chain $\ds \ra \dnull + \pi^+$
followed by 
$\dnull \ra K^- + \pi^+$. In total 35 $\pm$ 7.2 events could be observed resulting in a \ds yield in \nmu CC
interactions of
(0.79$\pm$0.17(stat.)$\pm$0.10(syst.)) \% \cite{nodstar}.

\section{Summary}
Both recent short baseline experiments at CERN, \ch and \no, boosted the limit on the mixing angle
\sint for \delm $>$ 10 \evs by an order of magnitude and more with respect to former experiments.
The improvement was obtained in the \nmuntau and \nentau channel as well. A recent \no analysis in the
\nmune channel can exclude the parameters of the LSND evidence for \delm $>$ 10 \evs. Several interesting
results on charm physics were already obtained and can be expected in the future.

\section*{Acknowledgments}
This work is supported by a Heisenberg Fellowship of the Deutsche Forschungsgemeinschaft. I would
like to thank my collegues from \ch and \no for valuable discussions.


\section*{References}
\begin{thebibliography}{99}
\bibitem{kai} K. Zuber, \Journal{\PRP}{305}{295}{1998}  
\bibitem{chorus}E. Eskut {\it et al}, \Journal{\NIMA}{401}{7}{1997}
\bibitem{nomad} J. Altegoer {\it et al}, \Journal{\NIMA}{404}{96}{1998}
\bibitem{chnutau}E. Eskut  {\it et al}, \Journal{\PLB}{497}{8}{2001}
\bibitem{nonutau}  P. Astier  {\it et al}, \Journal{\NPB}{611}{3}{2001}
\bibitem{slava} V. Valuev, Proc. Int. Europhysics Conf. on High Energy Physics, Budapest 2001
\bibitem{lsnd} A. Aguilar  {\it et al}, \Journal{\PRD}{64}{112007}{2001}
\bibitem{karmen} B. Armbruster {\it et al}, hep-ex/0203021
\bibitem{klaus} E. D. Church {\it et al}, hep-ex/0203023
\bibitem{noosdm} P. Astier {\it et al}, \Journal{\PLB}{486}{35}{2000}
\bibitem{chdo} A. Kayis-Topasku {\it et al}, \Journal{\PLB}{527}{173}{2002}
\bibitem{nodstar}  P. Astier {\it et al}, \Journal{\PLB}{526}{278}{2002}

\end{thebibliography}
\end{document}